\documentclass{osa-article}

\journal{oe}

\articletype{Research Article}

\newcommand{\MJ}[1]{\textcolor{black}{#1}}
\newcommand{\MJdwa}[1]{\textcolor{black}{#1}}

\usepackage{lineno}
\usepackage{esint}
\usepackage{amsmath}
\usepackage{graphicx}
\usepackage{epstopdf}
\usepackage{float}
\usepackage{siunitx}

\usepackage{color}
\definecolor{mygreen}{rgb}{0,0.5,0}
\definecolor{mygrey}{rgb}{0.5,0.5,0.5}
\definecolor{myred}{rgb}{0.75,0,0}
\definecolor{myblue}{rgb}{0,0,0.75}
\definecolor{mymagenta}{cmyk}{0,1,0,0.12}
\definecolor{mycyan}{cmyk}{1,0,0,0.12}
\definecolor{myorange}{rgb}{1.,0.5,0}
\definecolor{myviolet}{rgb}{0.6,0.15,0.6}
\definecolor{mybrown}{cmyk}{0,0.50,1,0.41}

\usepackage{ulem}

\global\long\def\et{\mathcal{E}}%
\global\long\def\fsym{\mathcal{F}}%
\global\long\def\infint#1{\intop_{-\infty}^{+\infty}\:\mathrm{d}#1}%
\global\long\def\ssym{\mathcal{S}}%
\global\long\def\ftd#1#2#3#4{\fsym_{\substack{#1\rightarrow#2\\
 #3\rightarrow#4 
}
 }}%
\begin{document}

\title{Variable electro-optic shearing interferometry for ultrafast single-photon-level pulse characterization}
\author{Stanisław Kurzyna,\authormark{1,2,*} Marcin Jastrzębski,\authormark{1,2,*}  Nicolas Fabre,\authormark{3} Wojciech Wasilewski,\authormark{1} Michał Lipka,\authormark{1,5} and Michał Parniak\authormark{1,4,6}}

\address{\authormark{1}Centre for Quantum Optical Technologies, Centre of New Technologies, University of Warsaw, Banacha 2c, 02-097 Warsaw, Poland\\
\authormark{2}Faculty of Physics, University of Warsaw, Pasteura 5, 02-093 Warsaw, Poland\\
\authormark{3}Departamento de Óptica, Facultad de Física, Universidad Complutense, 28040 Madrid, Spain\\
\authormark{4}Niels Bohr Institute, University of Copenhagen, Blegdamsvej 17, 2100 Copenhagen, Denmark\\
\authormark{5}m.lipka@cent.uw.edu.pl\\ 
\authormark{6}m.parniak@cent.uw.edu.pl\\} 
\address{\authormark{*}These authors contributed equally}


\begin{abstract}
Despite the multitude of available methods, the characterisation of ultrafast pulses remains a challenging endeavour, especially at the single-photon level.
We introduce a pulse characterisation scheme that maps the magnitude of its short-time Fourier transform. Contrary to many well-known solutions it does not require nonlinear effects and is therefore suitable for single-photon-level measurements. Our method is based on introducing a series of controlled time and frequency shifts, where the latter is performed via an electro-optic modulator allowing a fully-electronic experimental control. \MJ{We characterized the full spectral and temporal width of a classical and single-photon-level pulse and successfully tested the applicability of the reconstruction algorithm of the spectral phase and amplitude.} The method can be extended by implementing a phase-sensitive measurement and is naturally well-suited to partially-incoherent light. 
\end{abstract}
\section{Introduction}

Ultrafast optical pulse characterization remains a challenging endeavor
and generally requires scenario-specific techniques tailored to the
pulse central wavelength, bandwidth and power. Among possible approaches
non-linear spectrographic (such as FROG \cite{Kane1993}) and interferometric
(such as SPIDER \cite{Iaconis1999}) methods prevail. Interestingly,
the recent advent of high bandwidth electro-optical modulators (EOM)
and fiber Bragg grating (FBG) spectrometers enabled implementations
of highly-sensitive linear electro-optic shearing interferometry (EOSI)
\cite{Dorrer2003,Kang2003a}, first proposed by Wong and Walmsley
\cite{Wong1994} and recently demonstrated at the single-photon level
\cite{Davis2018,Davis2018a,davis_measuring_2020}. Being an interferometric technique,
EOSI avoids the caveat of spectrographic methods \textendash{} a large number
of measurement samples $\mathcal{O}(N^{2})$ required to reconstruct
$N$ points of the pulse's complex electric field. At the same time,
being a linear method it is compatible with the photon-starved regime,
where the low efficiency of frequency conversion limits non-linear
methods such as FROG or SPIDER including its variants \cite{Lelek2008,Birge2010}. 
 Nevertheless, EOSI still requires frequency-resolved detection, which can be a challenge for single photons. One solution is to use frequency-to-time mapping methods based on dispersion, for example in a long fiber.
This is the operating principle of single-photon spectrometers based on time-tagging photon counts with low-jitter detectors. 
More conveniently, FBGs are available at telecom wavelengths and allow such setups to be more compact. However, for the near-infrared band large dispersion is notoriously difficult to achieve. A traditional diffraction grating \cite{Lipka:21,Lipka2021prl} combined with a single-photon sensitive camera provides a good alternative. Nevertheless, in all cases the necessity of spectrally-resolved single-photon detection is the main factor that increases the cost and complexity of the setup.

In this work we experimentally demonstrate a new method of variable shearing interferometry (VarSI) which enables a measurement
of an ultrafast pulse spectrogram at the single-photon level. In our work the method operates
in the near-infrared band (\SI{800}{\nano\meter}) without a need for a single-photon spectrometer. \MJdwa{Similarly, Fourier transform chronometry presented in  \cite{golestani2022electrooptic} allows measuring the temporal width of the ultrashort single-photon pulses by sweeping spectral shifts.}
VarSI relies on a scan of both temporal delays and spectral shifts and
remains robust to interferometric phase fluctuations by exploiting a
phase-insensitive second-order measurement of intensity correlation, or simply in the classical regime a measurement of fringe visibility. The presented interferometric method not only \MJ{allows measuring} a spectrogram of classical fields, but the same optical configuration was initially proposed for measuring the chronocyclic Wigner function of the phase-matching function of frequency-entangled photon pairs \cite{douce_direct_2013}, and later experimentally realized \cite{tischler_measurement_2015}. However, in \cite{tischler_measurement_2015} the relative frequency shift of the photon pairs is performed during the generation process, either by varying the crystal temperature or by tuning the pump frequency. Here, the frequency shift is performed by using an EOM which provides perspectives for chip integration.

\section{Theory}

\subsection{Idea of VarSI} 
\begin{figure}[h]
\centering
\includegraphics[width=1\columnwidth]{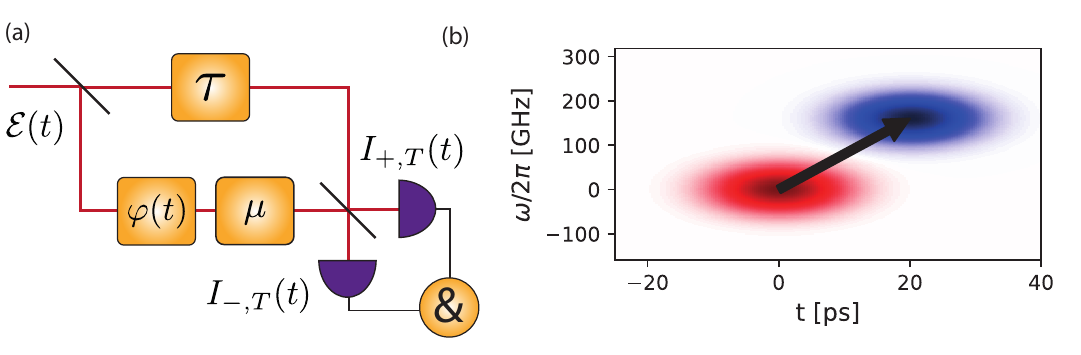}\caption{\label{fig:idea} 
(a) 
Variable shearing inteferometer (VarSI) with a second-order correlation or visibility-based detection.
On a balanced beamsplitter (BS) an input pulse with electric field
$\protect\et(t)$ is split into two replicas. One is delayed by $\tau$,
while the other spectrally shifted by $\mu$. The time-dependent phase
difference between the arms is incorporated into $\varphi(t)$. The
pulses are interfered on a second BS which output ports ($\pm$) are
observed with square-law photodetectors. Under fluctuating phase $\varphi(t)$
conditions, the second-order correlation function $g^{(2)}(\mu,\tau)$,
 or in the classical regime fringe visibility between the registered intensities, captures the self-gated Short Time
Fourier Transform (STFT) $|\protect\ssym_{\protect\et(t)}(\mu,\tau)|^{2}$
for a given pair of temporal and spectral shifts ($\mu,\tau$). (b) Ideational scheme of Wigner functions for a temporal and a spectral shift.}
\end{figure}

The idea of VarSI is depicted in Fig. \ref{fig:idea}. Similarly to other spectral shearing interferometry methods, the investigated pulse $\et(t)$ is split into two copies. One is delayed by $\tau$, yielding $\et(t-\tau)$, and the other shifted in frequency by $\mu$  which yields $\et(t)e^{i\mu t}$. Finally the pulses interfere at a balanced beamsplitter (BS) and exit through the $\pm$ ports. Compared to EOSI, in VarSI a spectrometer at the single output of
the spectral shearing interferometer is replaced with a pair of square-law
detectors observing both output ports. Importantly, the bandwidth
of the detectors can be very low \textendash{} the only requirement is the ability
to resolve temporal fluctuations of the phase difference between the interferometer
arms $\varphi(t)$. In particular, the essential role of the detectors is to average the signal at a much larger timescale $T$ \MJ{than} the duration of the pulse. 

\subsection{Classical case - analyzing fringe visibility}
The detectors produce a signal $I_{\pm}\equiv I_{\pm,T}(t)\propto\int_{t-T/2}^{t+T/2}\mathrm{d}t'\:|\et_{\pm}(t')|^{2}$
proportional to the optical intensity averaged over time $T$. 
In particular, for a given relative phase between arms $\varphi$, we observe intensities on both detectors:
\begin{equation}
I_{\pm} \propto \infint t\:|\et(t)|^{2} + \infint t\:|\et(t-\tau)|^{2} \pm 2\mathcal{V} \mathrm{Re}
\left(e^{i \varphi} \infint t\:\et(t)\et^*(t-\tau)e^{i\mu t}\right),
\label{eq:Ipm-fringe}
\end{equation} 
where $0\leq\mathcal{V}\leq1$ denotes the intrinsic interference visibility which may depend on various imperfections of the setup and for now shall be assumed $\mathcal{V}=1$.
The last term of Eq. (\ref{eq:Ipm-fringe}) contains information about the interference when the phase $\varphi$ is scanned over the entire $2\pi$ range. 

In particular, we note that it is related to the self-gated Short Time Fourier Transform (STFT):
\begin{equation}
\ssym_{\et(t)}(\mu,\tau)= \frac{1}{\sqrt{2\pi}} \infint t\et(t)\et^{\ast}(t-\tau)\exp(i\mu t).
\label{spectrotime}
\end{equation}
\MJ{Generally speaking we may consider incoherent states and its first order coherence function defined as:}
\begin{equation}
g^{(1)}(t,\tau) = \langle\et(t)\et^{\ast}(t-\tau)\rangle,
\label{general g1}
\end{equation}
\MJ{where $\langle \cdots \rangle$ is the statistical ensemble average and $g^{(1)}$ is then related to STFT in the following way:}
\begin{equation}
\ssym_{\et(t)}(\mu,\tau) = \frac{1}{\sqrt{2\pi}}\infint t\:g^{(1)}(t,\tau)e^{i\mu t}
\label{STFT and g1}
\end{equation}

\MJ{Therefore VarSI is a valid method of pulse characterization regardless of the coherence of light.} In VarSI the delay $\tau$ and spectral shift $\mu$ can be precisely scanned $\tau\in\lbrace\tau_j\rbrace_{j=1\ldots M}$, $\mu\in\lbrace\mu_k\rbrace_{k=1\ldots N}$.
For a given pair of shift $(\tau_j,\mu_k)$ and the phase $\varphi$, taking the difference between the output intensities $I^{(j,k)}_{+}-I^{(j,k)}_{-}$ separates the interference term. Averaging over the $\varphi$ scan and taking the standard deviation, we can effectively recover the visibility of the interference:
\begin{equation}
\sqrt{\langle [\mathrm{Re}(e^{i\varphi}\ssym_{\et(t)}(\mu_k,\tau_j))]^2\rangle_\varphi} = |\ssym_{\et(t)}(\mu_k,\tau_j)|^{2},
\end{equation}
which is equivalent to the modulus-squared STFT at $(\tau_j,\mu_k)$.
This way a series of $M\times N$ simple time-integrated non-resolving intensity measurements directly corresponds to mapping the modulus-squared STFT of the investigated pulse as the interference visibility (relative to a certain maximum visibility).

\subsection{Single-photon-level pulses}
Next, let us consider a case of intensities that are too weak for us to observe the fringes directly [which ability we assumed for Eq. (\ref{eq:Ipm-fringe})], and averaging is impossible due to phase instability. This could be, in our case, primarily due to rapid scanning of the time delay, which makes the phase stabilization difficult. In such case, to recover the visibility, we interrogate the second-order correlation.
Indeed, assuming the interferometer remains stable over the (relatively short)
averaging time $T$, for each ($\mu,\tau$), where here and henceforth we drop the $(j,k)$ indices, the product of intensities
can be averaged $\langle I_+(t)I_-(t)\rangle_{\varphi}$ to yield a second-order correlation which is (see \nameref{appendixA}):
\begin{equation}
    g^{(2)}(\mu,\tau)=\frac{\langle I_{+,T}(t;\mu,\tau)I_{-,T}(t;\mu,\tau)\rangle_{\varphi}}{\langle I_{+,T}(t;\mu,\tau)\rangle_{\varphi}\langle I_{-,T}(t;\mu,\tau)\rangle_{\varphi}}=
    1-\frac{1}{2}{\cal{V}}\frac{|\ssym_{\et(t)}(\mu,\tau)|^{2}}{(\infint t |{\cal{E}}(t)|^{2} )^2} ,
\label{g2}
\end{equation}
with the averages $\langle.\rangle_{\varphi}$ taken over a uniform
distribution of $\varphi$. 

\begin{figure}[H]
\centering
\includegraphics[width = 1\columnwidth]{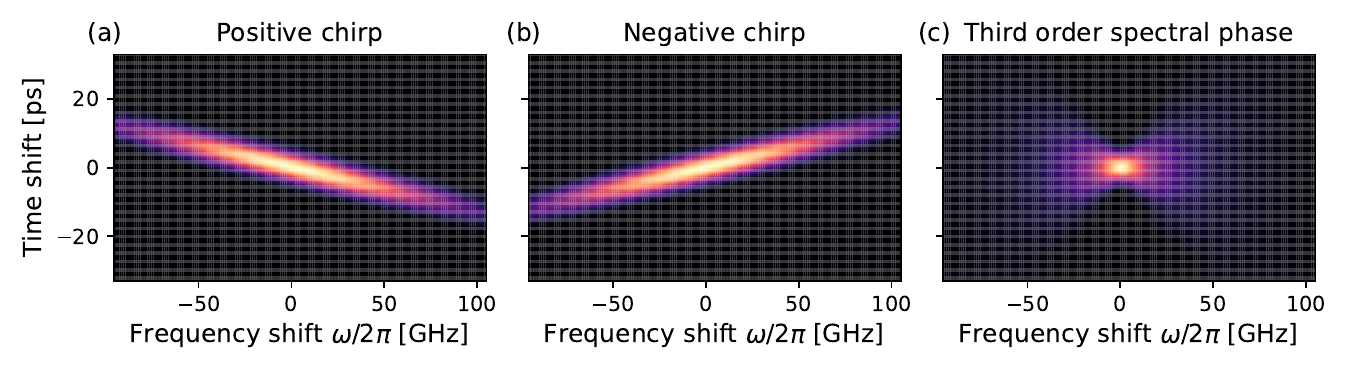}
\caption{\label{fig:chirp}Simulation of the modulus squared STFT $|\protect\ssym_{\protect\et(t)}(\mu,\tau)|^{2}$ of an exemplary Gaussian pulse with: (a) positive chirp, (b) negative chirp, (c) third order spectral phase}
\end{figure}

Note that the factor $1/2$ arises from the phase averaging and limits the maximal visibility for coherent states.
\paragraph{}
Exemplary spectrograms $|\ssym_{\et(t)}(\mu,\tau)|^{2}$ of a Gaussian pulse with
quadratic and cubic spectral phase components has been depicted
in Fig.~\ref{fig:chirp}. Note that this particular form of spectrograms allows unambiguous determination of the quadratic phase sign, similarly to the FROG technique. 
Finally, let us note that if the phase could be stabilized long-term, then the fringe measurements from Sec. 2.2 can still be applied for very weak single-photon level light, or for instance quantum states of light such as single photons.

\subsection{Relation to Chronocyclic Wigner Function}
An intricate and conceptually useful property of the STFT is that it directly relates to the Chronocyclic Wigner Function (CWF) of the characterized pulse.
Defining CWF as:
\begin{equation}
     W(\omega,t) = \frac{1}{2\pi}\infint t'\ \et\left(t + \frac{t'}{2}\right)\et^{*}\left(t - \frac{t'}{2}\right)e^{i\omega t'},
\end{equation}
and performing a two-dimensional Fourier transform, one obtains a relation between STFT and CWF:
\begin{equation}
    \ssym_{\et(t)}(\mu,\tau) =  \sqrt{2\pi}\ftd{t}{\mu}{\omega}{\tau}[W(\omega,t)]e^{-\frac{i\mu\tau}{2}}.
\end{equation}
Hence, the measured STFT modulus squared can be expressed as:
\begin{equation}
     |\ssym_{\et(t)}(\mu,\tau)|^{2} =  2\pi\ftd{t}{\mu}{\omega}{\tau}[W(\omega,t)] \ftd{t}{\mu}{\omega}{\tau}^{*}[W(\omega,t)],
\end{equation}
which, applying the convolution theorem, simplifies to:
 \begin{equation}
    |\ssym_{\et(t)}(\mu,\tau)|^{2} =  2\pi\ftd{t}{\mu}{\omega}{\tau}[W_{\et(t)}(\omega,t)*W_{\et(-t)}(\omega,t)].
\end{equation}
The modulus squared of the STFT corresponds to the two-dimensional Fourier transform of a convolution between the CWF of the original pulse and the CWF of a time-reversed one. Therefore, as depicted in Fig. \ref{fig:idea}(b), the procedure is to first make a convolution of the Wigner functions. Finally, we perform the two-dimensional Fourier transform to obtain the final result.

\subsection{The reconstruction of the pulse complex electric
field}
While the mapping from an electric field $\mathcal{M}:\et(t)\rightarrow|\ssym_{\et(t)}(\mu,\tau)|^{2}$
is not directly invertible, if we assume $\et(t)$ to be
a time-limited pulse, the ambiguities of $\mathcal{M}^{-1}:\MJ{|}\ssym_{\et(t)}(\mu,\tau)|^{2}\rightarrow\hat{\et}(t)$
are limited to the global phase, reflection and spectral or temporal shift.
\cite{Pinilla2022}. The latter two stem from the inability of
the method to infer the sign of the cubic part and linear spectral phase, respectively.

Retrieving $\et(t)$ is interestingly an inverse problem known to the radar remote sensing
technique with a range of available algorithm, well-understood ambiguities
and a subject of ongoing vivid research \cite{Pinilla2022}. 

Here, we adopt a practical approach and modify the COPRA phase retrieval algorithm recently developed by Geib et al. \cite{geib_common_2019} in order to reconstruct both the amplitude and phase of the pulse in the frequency domain. Indeed, the expression of the spectrogram measured in our optical interferometric scheme is quite different from the one obtained with non-linear interferometric schemes already analyzed in \cite{geib_common_2019}. The modification of the phase retrieval algorithm is detailed in the \nameref{appendixB}.

\section{Experimental setup}
\begin{figure}
\centering
\includegraphics[width = 0.7\columnwidth]{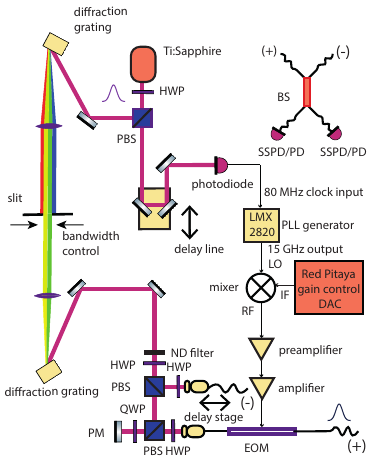}
\caption{\label{fig:setup} 
Experimental setup for VarSI. Pulses from a Ti:Sapphire laser are split on a polarizing beamsplitter (PBS). One part is sent through a delay line to adjust the slope of further generated RF signal. The pulse impinges onto a photodiode, resulting in an output waveform with a very sharp rising transition, which is used as a clock input of a PLL generator. The second part of the pulse is filtered using a 4f setup with an adjustable slit controlling the bandwidth of the pulse. The filtered pulse is attenuated (to the single-photon or $\sim$ 100$\mu$W level, depending on the scenario) by an neutral-density (ND) filter, split on a PBS and coupled to either a single mode polarization-maintaining fiber via a collimator on a delay stage $(+)$ or to a fiber connected to an electro-optic phase modulator (EOM) $(-)$. \MJ{Prior to EOM the light is reflected from scanning piezoelectric mirror (PM).} Half-wave plates (HWP) rotate the polarization to match the axes of the polarization-maintaining fibers and the crystal axis of the EOM. Resulting pulses $(\pm)$ are interfered on a fiber-based balanced BS which outputs are sent to superconducting single-photon detectors (SSPDs) or photodiodes (PD) for single-photon-level and classical scanario, respectively.
}
\end{figure}

The experimental setup for VarSI is depicted in Fig. \ref{fig:setup}. 
To prepare a test pulse, we use a 4f spectral filter ($\approx0.1\;\mathrm{nm}$
FWHM, 795$\;\mathrm{nm}$ central wavelength) and an attenuator on a $100\;\mathrm{fs}$
pulse from a Ti:Sapphire laser (SpectraPhysics MaiTai). 
For the VarSI itself, the pulse is split into two copies on a polarizing beamsplitter (PBS) and enters a fiber part of the
interferometer with one of the collimators placed on a motorized delay
stage (implementing temporal delay $\tau$). A spectral shift $\mu$ is obtained via electro-optic phase
modulation (EOM) synchronous with the pulses. 

\begin{figure}
\centering
\includegraphics[width = 0.9\columnwidth]{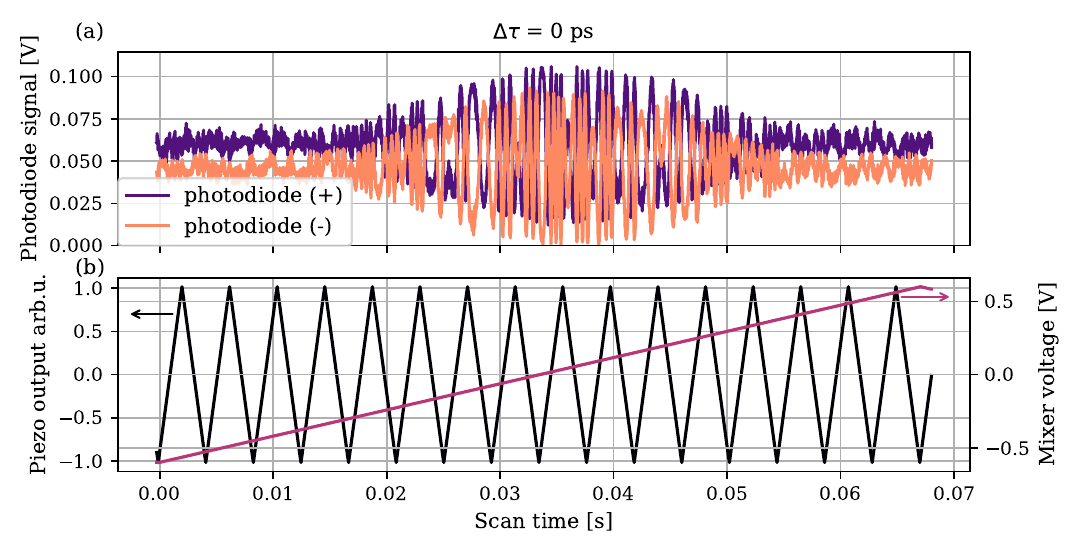}
\caption{\label{fig:oscilloscope} \MJ{(a) Signal from $\pm$ photodiodes (custom amplified photodectors) with interference fringes visible.
Temporal delay $\Delta \tau$ = 0 while spectral delay is varying (mixer scan),
(b) Piezo control signal (black) - is introduced to average the phase over piezo scans and frequency mixer IF signal (magenta)  -responsible for frequency shifts. The horizontal axis is common for both a) and b).}
}
\end{figure}

Two arms of the interferometer enter
a balanced fiber BS. The
signals at BS's output ports are observed either with photodiodes 
or again fiber coupled and sent to SSPDs (superconducting single-photon detectors - idQuantique ID281).
In case of photodiodes the signal exceeds 100 $\mu\mathrm{W}$ and the measurement of $\left|\mathcal{S}_{\mathcal{E}(t)}\right|^{2}$ is performed by taking the RMS of the fringe visibility in one period of piezo oscillation, which by itself contains several fringes, as depicted in Fig. \ref{fig:oscilloscope}.

For the measurement at the single-photon level the setup is modified by introducing ND filters before the SSPDs. The average number of photons per pulse after the interferometer and detector
losses is ca. $\bar{n}= 0.013$. 
Conveniently, fine tuning of the setup can be done by monitoring the number of single counts and coincidences at the SSPDs.

In all cases the measurements are collected by first scanning over the frequency shifts (electronic control signal) for each delay (motorized movement of the translation stage).

\subsection{EOM and RF setup}

To produce a spectral shear we drive the EOM synchronously with the laser pulse repetition rate using a phase-locked loop (PLL) generator. A \SI{30}{\milli\watt} average power of the beam is picked-off with a PBS, undergoes a delay with a custom-made free-space motorized delay line and is focused onto a photodiode (PD). The signal from the PD with ca. \SI{2}{\volt} peak-to-peak is sent into the clock input of a PLL generator based on the LMX2820 chip (Texas Instruments). The quality of the phase lock is essential to obtain a stable spectral shear. The PLL is programmed to output a $\sim$ \SI{15}{\giga\hertz} (or exactly 184 $\times$ \SI{80}{\mega\hertz} = \SI{14720}{\giga\hertz}) signal with \SI{-2.3}{dBm} of power. The signal is sent to a mixer (Mini-Circuits ZX05-24MH-S+) input (LO) in order to allow control of its amplitude with a level of the DC voltage on the IF mixer input. Note that the IF mixer input must be DC-coupled. In such case, care must be taken to avoid overvoltage in order to protect the sensitive RF transformers and diodes of the mixer. The mixer output is amplified using two subsequent amplifiers (Mini-Circuits ZX60-06183LN+ and ZVE-3W-183+) and then directed to the EOM. In this configuration we achieve mean power of the \SI{15}{\giga\hertz} sine wave of \SI{28}{dBm}. The DC control is achieved using a Red Pitaya STEMlab 125-14 programmable logic board's 14-bit DAC output with a \SI{\pm1}{\volt} range. Notably, in general previous approaches to varying the signal amplitude at the EOM for controllable spectral shearing or temporal lensing relied on generating different signals at their source \cite{PhysRevApplied.14.014052}.

With the optical pulse traveling through the EOM much quicker than
the inverse of the RF driving frequency, we can approximate the amount
of the frequency shift (in radians per second) imposed by the EOM as:
\begin{equation}
\mu=2\pi\times\frac{V_{\mathrm{pp}}f_{\mathrm{RF}}}{2V_{\pi}},
\label{eq:eommu}
\end{equation}
where we have assumed the pulse is temporally aligned with the steepest part
of the RF signal and where $V_{\mathrm{pp}}$ denotes the peak-to-peak
driving voltage at the EOM, $f_{\mathrm{RF}}$ is the signal's fundamental
frequency and $V_{\pi}$ corresponds to the EOM's driving voltage
required for a $\pi$ phase shift. In our case $V_{\pi}=4/\pi\;\mathrm{V}/\mathrm{rad}$,
$f_{\mathrm{RF}}\approx15\;\mathrm{GHz}$ giving $\mu/V_{\mathrm{pp}}=6\;\mathrm{GHz}/\mathrm{V}$.

\subsection{Spectral shear calibration}

\begin{figure}[b]
\centering
\includegraphics[width = 1\columnwidth]{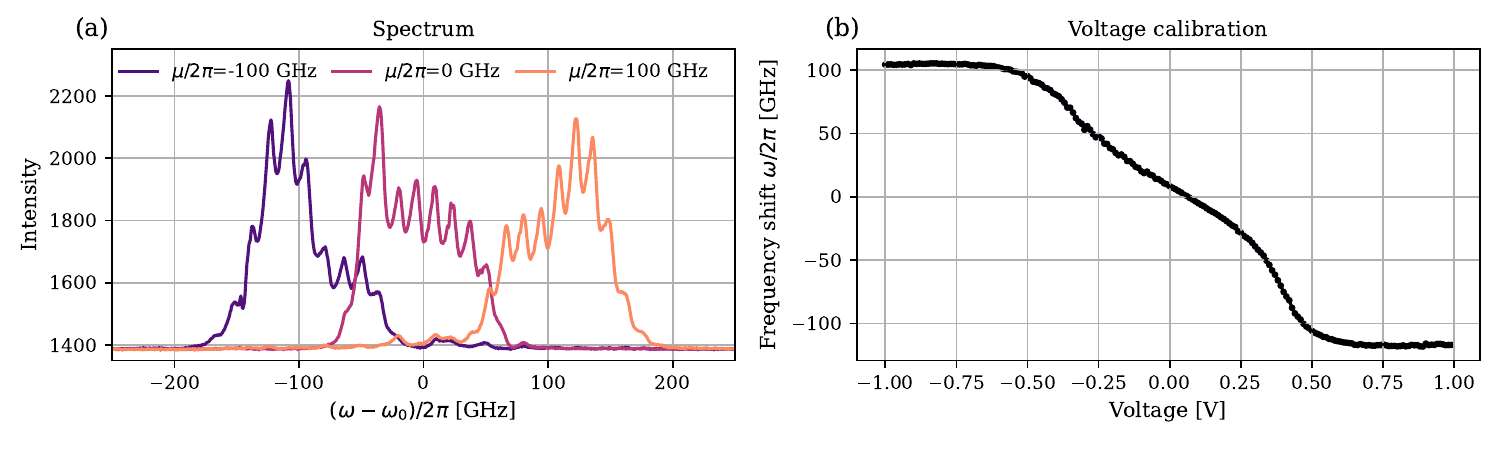}
\caption{\label{fig:calibration}
(a) Calibration spectra collected with a custom spectrometer for a series of frequency shifts from \SI{-100}{\giga\hertz} to \SI{100}{\giga\hertz}. 
(b) Frequency shift estimated from the spectral centroid for subsequent values of the frequency-shifter control signal.
}
\end{figure}

An important element of the VarSI calibration is establishing a relationship between the control signal and the resulting frequency shift. The control signal is scanned between $\SI{-1}{\volt}$ and $\SI{1}{\volt}$ while a custom spectrometer provides an estimate of the frequency-shifted pulse's spectral centroid. A calibration curve is obtained by performing a linear interpolation on the obtained data points and allows fine control of the frequency shifter. Exemplary calibration results are depicted in Fig. \ref{fig:calibration}.

For spectral shift calibrations we use a custom-built grating (1200
ln/mm, 750 nm blaze) spectrometer in a double-pass second-order configuration.
 The spectrometer utilizes a line-camera (Toshiba TCD1304AP) controlled with an STM32f103c8t6
microcontroller (``bluepill'' board). During its operation line images
from the sensor are sent to PC and averaged. Finally, a Gaussian curve is fitted
yielding its center as the spectrometer readout.

The spectrometer has been calibrated with an external adjustable continous-wave (CW) external-cavity diode laser (ECDL)
(Toptica DL100) and a High Finesse WS/6 wavemeter by collecting a series of $360$
measurements between \SI{794.2}{\nano\meter} and \SI{795.99}{\nano\meter}. A polynomial fit to the
gathered calibration data serves as a calibration curve for further
measurements.

The spectrometer facilitates the delay matching between the RF-driving
optical pulse and the experiment as well as calibration of the spectral
shift as a function of the mixer voltage. In practice, we align the relative delay by maximizing the spectral shift observed on the spectrometer. At other non-optimal delays we also observe strong distortions of the spectrum due to time-lensing effects. Finally, we note that even at the optimal point the spectrum is distorted, as the tails of the pulse stretch to the times outside of the linear voltage slope. This means that we may not be able to properly interfere and thus identify and reconstruct parts of the pulse that are very far in time from the time-axis origin of the STFT maps.

\section{Results}
\begin{figure}[h]
\centering
\includegraphics[width = 1\columnwidth]{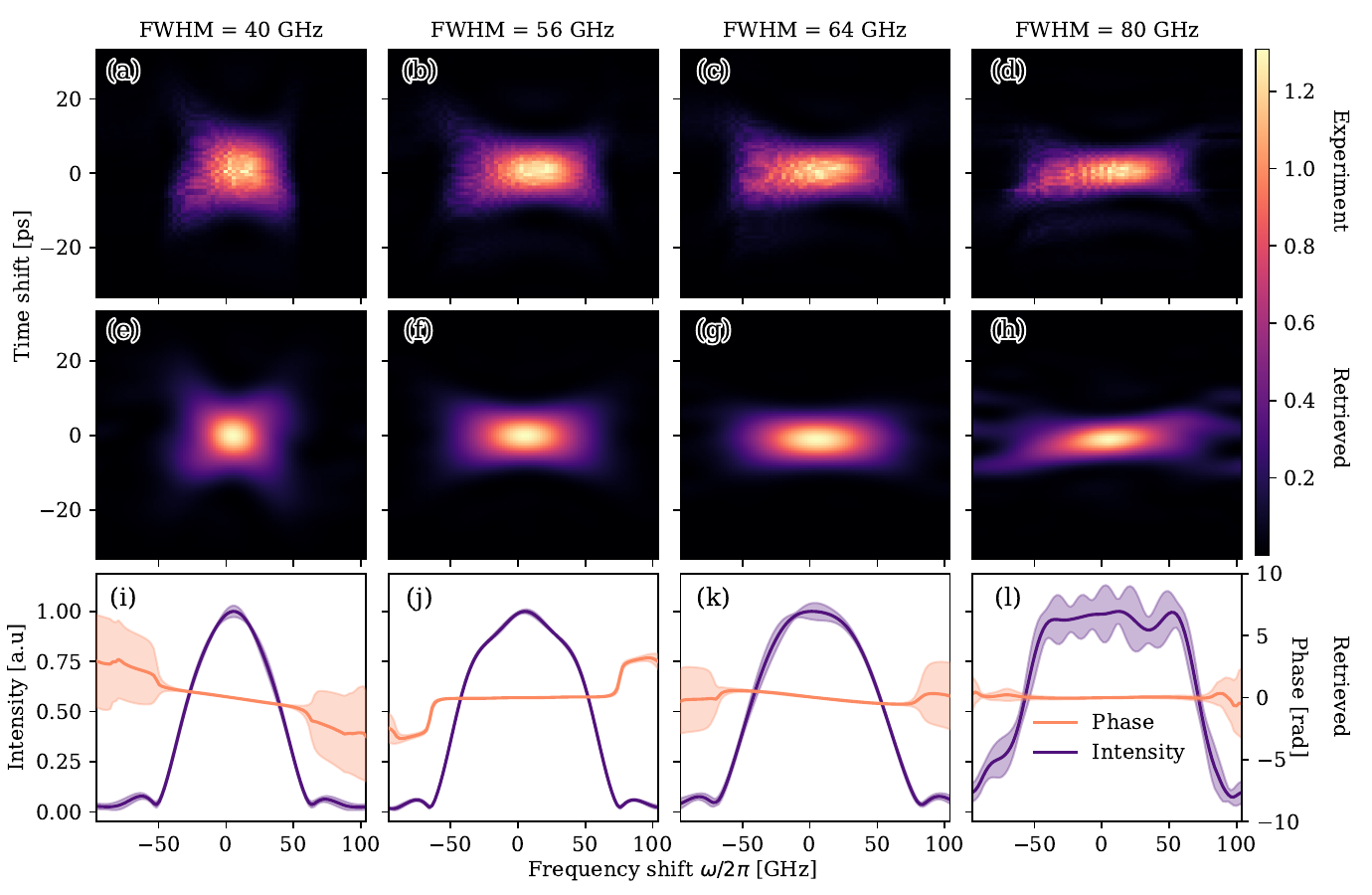}
\caption{\label{fig:classical} (a)-(d) Experimental spectrograms for classical-regime pulses with bandwidths between $\SI{48}{\giga\hertz}$ (first column) and $\SI{96}{\giga\hertz}$ (last column). (i)-(l) Corresponding reconstructed complex electric field (magnitude and phase) in the spectral domain with error bars. (e)-(f) Spectrograms calculated from the reconstructed pulses. 
}
\end{figure}

\begin{figure}[ht]
\centering
\includegraphics[width = 1\columnwidth]{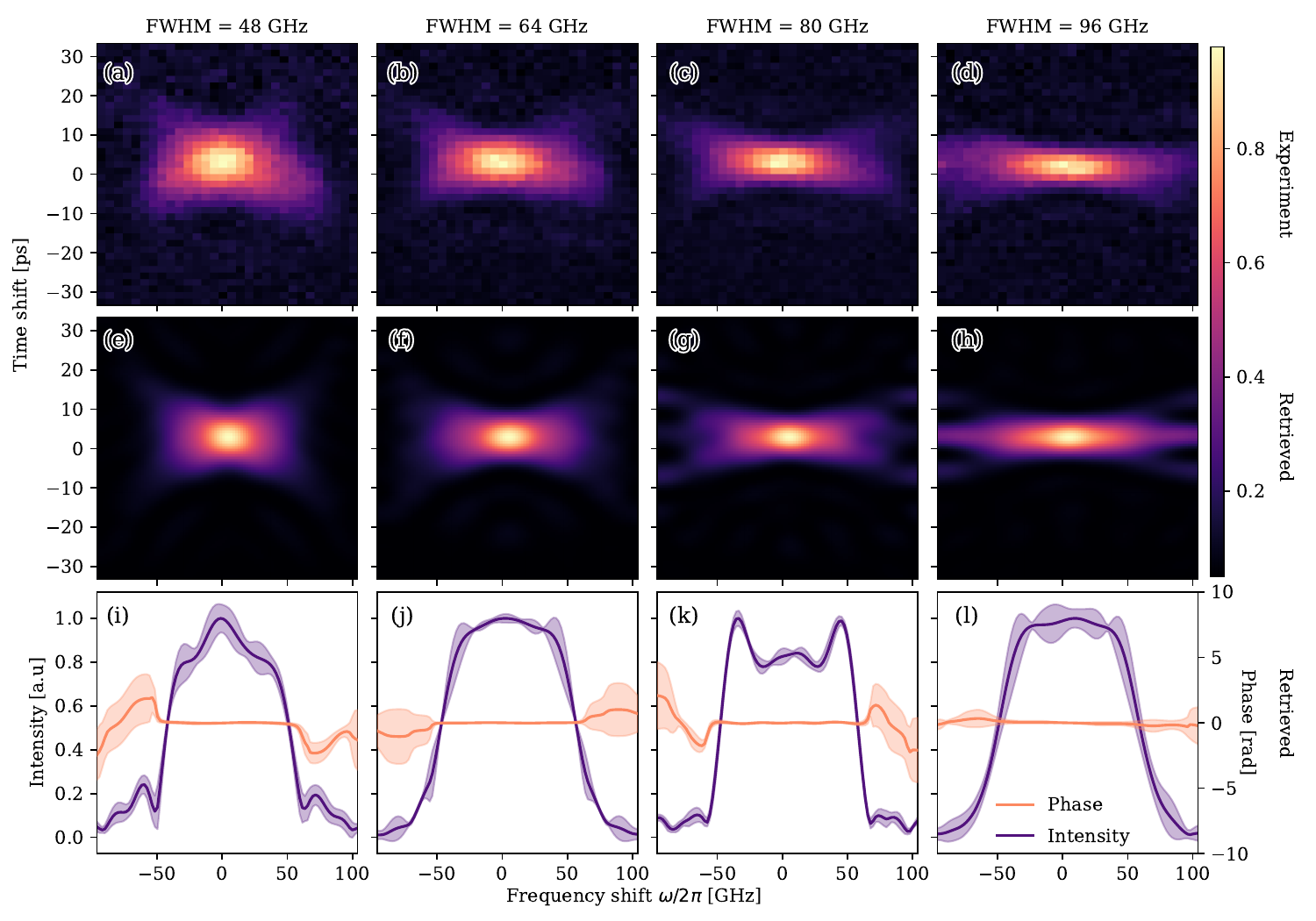}
\caption{\label{fig:results} 
 (a)-(d) Experimental spectrograms for single-photon-level pulses with bandwidths between $\SI{48}{\giga\hertz}$ (first column) and $\SI{96}{\giga\hertz}$ (last column). (i)-(l) Corresponding reconstructed complex electric field (magnitude and phase) in the spectral domain with error bars. (e)-(f) Spectrograms calculated from the reconstructed pulses. 
}
\end{figure}

\begin{figure}[ht]
\centering
\includegraphics[width = 1\columnwidth]{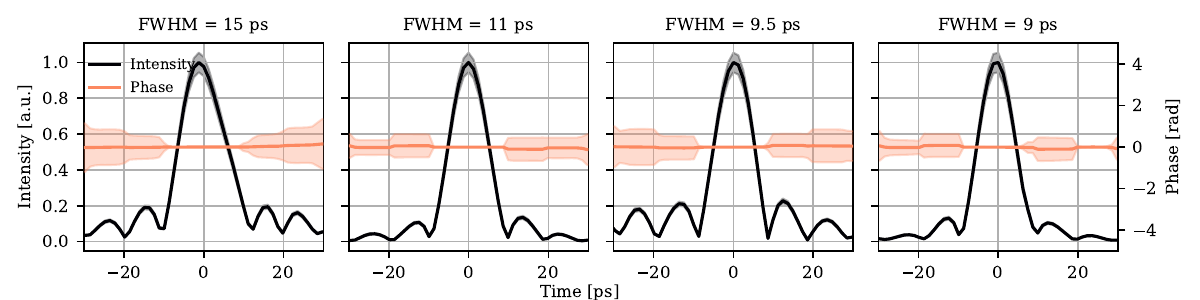}
\caption{\label{fig:sinc} Fourier transform of the reconstructed electric field for single-photon-level pulses with error bars.
}
\end{figure}

We have demonstrated the VarSI method in the near-infrared regime ($\sim$\SI{800}{\nano\meter}) for ultrafast (\SI{100}{\fs}) pulses filtered to selected bandwidths between \SI{48}{\giga\hertz} and \SI{96}{\giga\hertz} with both relatively high power pulses as well as single-photon-level light. By designing a non-Gaussian spectral filtering, we were able to shape and observe fine features of the pulses' spectrograms, proving the method's resolving capabilities. 
As depicted in Fig. \ref{fig:classical} for classical-regime pulses and in Fig. \ref{fig:results} at the single-photon level, the experimental data closely matches spectrograms of the reconstructed pulses. For reconstruction, the spectrograms of measured data were linearly interpolated. The bottom-most rows of Fig. \ref{fig:classical} and Fig. \ref{fig:results} depicts the amplitude and phase of the reconstructed pulses in the spectral domain, while their counterparts in the temporal domain are depicted in Fig. \ref{fig:sinc}. \MJ{The reconstructed phases are flat as expected which proves the correctness of the algorithm.}

The pulse retrieval quality of VarSI can be quantified with the fidelity
\begin{equation}
\mathrm{F} =\frac{\langle |\mathcal{S}_\mathrm{exp}| |\mathcal{S}_\mathrm{recon}| \rangle}{\sqrt{\langle|\mathcal{S}_\mathrm{exp}|^2 \rangle  \langle |\mathcal{S}_\mathrm{recon}|^2\rangle}}, 
\end{equation}
where the average $\langle . \rangle$ is taken over the extent of the experimental $|\mathcal{S}_\mathrm{exp}|^2$ and reconstructed $|\mathcal{S}_\mathrm{recon}|^2$ spectrograms. \MJ{Note that in this case fidelity quantifies how well the impulse is reconstructed in post-processing and not in respect to some ground truth.} In the classical regime we obtain $\mathrm{F}_\mathrm{class}=94\%$ while at the single-photon level $\mathrm{F}_\mathrm{sp}= 97\%$. We note that the retrieval procedure was interrupted after 300 iterations. 

Finally, let us discuss the limitations of the presented method. Narrowing the pulse spectrally causes expansion of temporal width which may lead to the pulse being unsuitable for the frequency-shifting sine waveform, as it will no longer fit within the linear slope. 
\MJ{Another limitation is the available range of amplitude and phase of the pulse. Most importantly the maximum optical input must not exceed the maximal optical input power of EOM. The available phase range also has its potential limitations e.g. finite resolution of spectral phase features present in the pulse.} 
On the other hand, if the bandwidth is too large then frequency shifts generated by EOM may not exceed it. In such a case, the entire map cannot be observed, and only limited information about the pulse can be retrieved. 
Notably, even though the frequency shift could be enhanced via modulation at a higher RF frequency [Eq. (\ref{eq:eommu})], it would in turn tighten the restriction on the pulse length. In order to extend the range of suitable bandwidth an EOM with lower $V_{\pi}$ as in \cite{Liu:21} ($1.75\ \mathrm{V}$) or \cite{Wang2018} ($1.4\ \mathrm{V}$), for example via developments in thin-film modulators \cite{zhu2021spectral}
would be required. This indeed is an active area of experimental development.
An interesting alternative approach would be to use a serrodyne waveform \cite{Johnson:10}, which for appropriate power and frequency range would be a challenge in electronic engineering. Furthermore, one would have to tune the frequency rather than the amplitude.

\section{Conclusions}
We have demonstrated a novel method - Variable Shearing Interferometry (VarSI) - for ultrafast optical pulse characterization at the single-photon level. The technique, while being a variant of shearing interferometry, avoids inefficient non-linear frequency conversion (such as commonly used in e.g. FROG or SPIDER), and does not require a spectrally-resolved measurement which remains challenging to implement for near-infrared single-photon signals.
Employing a state-of-the-art reconstruction algorithm we retrieved a complex electric field for a series of test pulses at $\SI{800}{\nano\meter}$ with bandwidths ranging from $\SI{48}{\giga\hertz}$ to $\SI{96}{\giga\hertz}$ and non-Gaussian spectral profiles, achieving high reconstruction fidelity $\mathrm{F}_\mathrm{sp}$ = 97\% ($\mathrm{F}_\mathrm{class}=94\%$).
\MJ{To calculate errors of reconstruction we repeated the algorithm many times with random starting points. For each spectral point, a histogram of reconstructions gives the mean and a standard deviation – the error bar.}
Importantly, even without any reconstruction, VarSI directly measures the modulus squared of a Short Time Fourier Transform (STFT) of the pulse \textendash{} a time-frequency distribution commonly employed in signal processing and yielding a plethora of information itself e.g. on the amount of chirp in the investigated pulse. Our method is also applicable to pulses with partial spectral coherence.
Notably, VarSI can be extended to a phase-sensitive implementation allowing direct measurement of a complex STFT (containing complete information on the pulse electric field) by incorporating a phase-tracking pilot-wave into the interferometer. Finally, the method described in this paper can also be applied for performing the spectral tomography of a single photon using another fully characterized one \cite{fabre_spectral_2022}. 
With the rapid development of quantum optical technologies and the growing interest in the spectral-temporal domain, a precise and versatile characterization of ultrafast, spectrally-broadband pulses at the single-photon level remains a timely topic. VarSI constitutes a novel, alternative solution to this standing problem, which fully utilizes recent technological developments in electro-optic modulation and circumvents the inaccessibility of single-photon-level spectrally-resolved detection for selected wavelengths. 
We also believe it goes significantly beyond the most recent state-of-the-art demonstration \cite{golestani2022electrooptic}, as the measurement of both time and frequency shifts (as opposed to only frequency shifts) \MJdwa{contains information about both amplitude and phase and allows reconstruction with only some ambiguities.} Furthermore, the method provides an experimental outlook into fundamentally important time-frequency phase-space of ultrafast pulses. Finally, with numerous prominent extensions, such as an on-chip integration or a phase-sensitive variant, we envisage prominent applications and a vast horizon of further research, especially in the context of temporal modes \cite{PhysRevX.5.041017} and super-resolution in time and frequency \cite{Mazelanik2022,PhysRevLett.121.090501}.

\section*{Appendix A}\label{appendixA}
\subsection*{Derivation of the $g^{(2)}$ function}
In this section, we derive the $g^{(2)}$ function measured with the electro-optic shearing interferometer from the theoretical results presented in \cite{sadana_near-100_2019,legero_characterization_2006}. We consider a pulse of linearly polarized light traveling along a given axis $z$ noted ${\cal{E}}(t)$. Such a pulse is split into two spatial paths noted $1,2$ with a balanced beam-splitter. The output field in the spatial mode $1$ and $2$ can be written as ${\cal{E}}_{1,2}(t)={\cal{E}}(t)/\sqrt{2}$.
A relative frequency and time shift are performed into the spatial paths $1$ and $2$. The electric field in each path is transformed as:
\begin{align}
{\cal{E}}_{1}(t)\rightarrow& \frac{{\cal{E}}(t-\tau)}{\sqrt{2}},\\
{\cal{E}}_{2}(t)\rightarrow& \frac{{\cal{E}}(t)e^{i\mu t}e^{i\varphi(t)}}{\sqrt{2}}.
\end{align}
The fluctuation of the input field is modeled by a random variable $\varphi$, whose probability distribution $P$ obeys:
\begin{equation}\label{definitionproba}
\int d\varphi P(\varphi) \text{cos}(\varphi)=0.
\end{equation}
The spatial paths intersect at another balanced beam-splitter, and the output spatial paths will be noted $\pm$. The output fields after the second beam-splitter can be written as:
\begin{equation}
{\cal{E}}_{\pm}(t)=\frac{1}{2}\left({\cal{E}}(t-\tau)\pm {\cal{E}}(t)e^{i\mu t}e^{i\varphi(t)}\right).
\end{equation}
Non-resolved time intensity detection is then measured in each arm of the interferometer, and the integration range will be noted $T$. The measured intensity can then be written as:
\begin{equation}
\int_{-T/2}^{T/2} \left|{\cal{E}}_{\pm}(t)\right|^{2} dt= \frac{1}{4} \left[\int_{-T/2}^{T/2} dt \left(|{\cal{E}}(t-\tau)|^{2} + |{\cal{E}}(t)|^{2} \pm 2 \text{Re}\left({\cal{E}}(t){\cal{E}}^{*}(t-\tau)e^{i\mu t}e^{i\varphi(t)}\right)\right)\right].
\end{equation}
We consider the integration range $T$ to be larger than the width of the pulse and can be extended to the infinity, which means that the first two terms are equals. The cross correlation, or the $g^{(2)}$ function, can be written as:
\begin{equation}
g^{(2)}(\tau,\mu)=\frac{\int d\varphi P(\varphi)\infint t |{\cal{E}}_{+}(t)|^{2} \infint t' |{\cal{E}}_{-}(t')|^{2} }{(\int d\varphi P(\varphi)\int_{-\infty}^{+\infty} dt |{\cal{E}}_{+}(t)|^{2}) (\int d\varphi P(\varphi)\int_{-\infty}^{+\infty} dt |{\cal{E}}_{-}(t)|^{2})},
\end{equation}
where we have averaged over the phase fluctuation. The denominator is equal to 
\begin{equation}
\left(\int d\varphi P(\varphi)\infint t |{\cal{E}}_{+}(t)|^{2}\right)\left(\int d\varphi P(\varphi)\infint t |{\cal{E}}_{-}(t)|^{2}\right)=\frac{1}{4}\left(\infint t |{\cal{E}}(t)|^{2} \right)^2.
\end{equation}
To prove this, we observed that 
\begin{multline}
\int d\phi P(\phi) \infint t \text{Re}({\cal{E}}(t){\cal{E}}^{*}(t-\tau)e^{i\mu t}e^{i\varphi(t)})=\int d\phi P(\phi) \infint t |{\cal{E}}(t){\cal{E}}^{*}(t-\tau)|\\
\times \text{cos}(\phi_{K}(t,\tau,\mu)+\phi(t))=0
\end{multline}
where we can directly performed the integral over the phase and we used Eq.~(\ref{definitionproba}) and the notation $\phi_{K}(t,\tau,\mu)=\text{arg}({\cal{E}}(t){\cal{E}}^{*}(t-\tau)e^{i\mu t})$. The numerator of the $g^{(2)}$ function is:
\begin{equation}
\frac{1}{4} \left[\left(\infint t |{\cal{E}}(t)|^{2}\right)^{2}-\int P(\varphi) d\varphi \left(\infint t \text{Re}\left({\cal{E}}(t){\cal{E}}^{*}(t-\tau)e^{i\mu t} e^{i\varphi(t)}\right)\right)^{2}\right].
\end{equation}
We have used Eq.~(\ref{definitionproba}), and we recognized the short-time Fourier transform ${\cal{S}}_{{\cal{E}}(t)}(\tau,\mu)$ defined by Eq.~(\ref{spectrotime}) of the field ${\cal{E}}$ and the window function corresponds to the conjugated field ${\cal{E^{*}}}$. The phase fluctuations are considered constant over the time where the two fields $E(t)$ and $E^{*}(t-\tau)$ are overlapping. Thus, we obtain for the second term of the numerator:
\begin{multline}
\int P(\varphi) d\varphi \left(\infint \text{Re}\left({\cal{E}}(t){\cal{E}}^{*}(t-\tau)e^{i\mu t} e^{i\varphi(t)}\right)\right)^{2}=\int P(\varphi) d\varphi \left(\text{Re}\left(e^{i\varphi} {\cal{S}}_{{\cal{E}}(t)}(\tau,\mu)\right)\right)^{2}\\
= \int P(\varphi)d\varphi( \text{cos}^{2}(\varphi) (\text{Re}({\cal{S}}_{{\cal{E}}(t)}(\tau,\mu))^{2}+\text{sin}^{2}(\varphi) (\text{Im}({\cal{S}}_{{\cal{E}}(t)}(\tau,\mu))^{2}\\
-2\text{cos}(\varphi)\text{sin}(\varphi) \text{Re}({\cal{S}}_{{\cal{E}}(t)}(\tau,\mu))\text{Im}({\cal{S}}_{{\cal{E}}(t)}(\tau,\mu)))
\end{multline}
Using again the property Eq.~(\ref{definitionproba}), and the fact that $\int d\varphi P(\varphi)  \text{cos}^{2}(\varphi)=\int d\varphi P(\varphi) \text{sin}^{2}(\varphi)$, the $g^{(2)}$ function yields,
\begin{equation}
g^{(2)}(\tau,\mu)=1-\frac{|{\cal{S}}_{{\cal{E}}(t)}(\tau,\mu)|^{2}}{(\int_{0}^{\infty} |{\cal{E}}(t)|^{2} dt)^2} \int d\varphi P(\varphi) \text{cos}^{2}(\varphi).
\end{equation}
As we are in the case where the phase is not stabilized, the integral over the phase can be simplified:
\begin{equation}
\int d\varphi P(\varphi) \frac{1}{2}(1+\text{cos}(2\varphi))= \frac{1}{2}.
\end{equation}
We finally arrive to the final result:
\begin{equation}
g^{(2)}(\tau,\mu)=1-\frac{1}{2} \frac{|{\cal{S}}_{{\cal{E}}(t)}(\tau,\mu)|^{2}}{(\int_{-\infty}^{+\infty} |{\cal{E}}(t)|^{2} dt)^2} 
\end{equation}
corresponding to Eq.~(\ref{g2}).

\section*{Appendix B}\label{appendixB}
\subsection*{Derivation of the gradient function and numerical errors}
We employ the notation in \cite{geib_common_2019} for deriving the expression of the gradient used in the common pulse retrieval algorithm (COPRA). The discretized electric field in the time and frequency domain are related by a Fourier transform:
\begin{align}
    \tilde{E}_{n}=& \sum_{n} D_{nk} E_{k}= \text{FT}_{n\rightarrow k}(E_{k})  \\
    E_{k}=& \sum_{k} D^{-1}_{kn} \tilde{E}_{n}, 
\end{align}
where $D_{nk}=\frac{\Delta t}{2\pi}e^{i\omega_{n}t_{k}}$ and $D^{-1}_{kn}=\Delta \omega e^{-i\omega_{n}t_{k}}$. The shifted pulse will be noted $A_{mk}=FT^{-1}_{n\rightarrow k}(e^{i\tau_{m}\omega_{n}}\tilde{E}_{n})$. We define the signal as: $S_{mk}=A_{mk}E^{*}_{k}$. 
The iterative algorithm for retrieving the phase and amplitude of the electric field from the measured spectrogram is summarized as follows. We start from an initial guess of the electric field $E$, and calculated its associated guess signal $S_{mn}$. A projection of the measured intensity is performed, which allows defining $S'_{mn}$ with the measured spectrogram and the guess signal (see Eq.~(14) in \cite{geib_common_2019}). The distance between the discretized signal and its projection is noted $Z_{m}$. Then, we evaluate the new electric as $E'_{n}=E_{n}-\gamma_{m} \Delta_{n} Z_{m}$, where $\gamma$ is the step size related to the convergence of the local iteration. $E'_{n}$ becomes the new guess value of the iterative algorithm. The algorithm is interrupted after 300 iterations. 

Importantly, we give the expression of the gradient of $Z_{m}$ related to the measured spectrogram specific to our experimental set-up (see Eq.~(\ref{spectrotime})), which takes the form
\begin{equation}
\Delta_{n} Z_{m}=-2\sum_{k} \Delta S^{*}_{mk} \frac{\partial S_{mk}}{\partial \tilde{E}^{*}_{n}}+\Delta S_{mk} \left[\frac{\partial S_{mk}}{\partial \tilde{E}_{n}}\right]^{*}.
\end{equation}
where the derivative of the STFT is:
\begin{equation}
\frac{\partial S_{mk}}{\partial \tilde{E}_{n}}= D^{-1}_{kn} e^{i\tau_{m}\omega_{n}} E^{*}_{k}+D^{*-1}_{kn} A_{mk}.
\end{equation}
Since in our case there is no dependence on the conjugated field, thus:
\begin{equation}
\Delta_{n} Z_{m}=-2\sum_{k} \Delta S_{mk} [D^{*-1}_{kn} e^{-i\tau_{m}\omega_{n}} E_{k}+D^{-1}_{kn}A^{*}_{mk}].
\end{equation}
The final expression of the gradient of $Z_{m}$ is:
\begin{equation}
\Delta_{n} Z_{m}=-\frac{4\pi \Delta\omega}{\Delta t} (\text{FT}_{n\rightarrow k}(\Delta S_{mk} E_{k})+\text{FT}^{*}_{n\rightarrow k}(\Delta S^{*}_{mk} A_{mk})).
\end{equation}

\begin{backmatter}
\bmsection{Funding}
Fundacja na rzecz Nauki Polskiej (MAB/2018/4 “Quantum Optical Technologies”); European Regional Development Fund; Narodowe Centrum Nauki (2021/41/N/ST2/02926); Ministerstwo Edukacji i Nauki (DI2018 010848).
\bmsection{Acknowledgments}
The “Quantum Optical Technologies” project is
carried out within the International Research Agendas programme of the
Foundation for Polish Science co-financed by the European Union under the
European Regional Development Fund.
This research was funded in whole  or in part by National Science Centre, Poland 2021/41/N/ST2/02926. For the purpose of Open Access, the author has applied a CC-BY public copyright license to any Author Accepted Manuscript (AAM) version arising from this submission. ML was supported by the Foundation for Polish Science (FNP) via the START scholarship.
This scientific work has been funded in whole or in part by Polish science budget funds for years 2019 to 2023 as a research project within the “Diamentowy Grant” programme of the Ministry of Education and Science (DI2018 010848). The phase retrieval algorithm which has been used, called pypret written by Niels C. Geib and available from GitHub, is licensed under the MIT license. SK and MJ contributed equally to this work.
We also wish to thank K. Banaszek and M. Mazelanik for the support and discussions.

\bmsection{Disclosures}
SK, MJ, WW, ML, MP are co-authors of a related patent application (P); The authors declare no other conflicts of interest.

\bmsection{Data availability} Data for figures 4-8 has been deposited at \cite{Data22} (Harvard Dataverse).


\end{backmatter}


\bibliography{refs}

\end{document}